\documentclass[
 twocolumn,
 amssymb,amsmath,9pt]{revtex4}[10pt]
\usepackage{graphicx}
\usepackage{dcolumn}
\usepackage{bm}
\usepackage{amssymb}
\usepackage{epsfig}
\newcommand{\be}{\begin{equation}}
\newcommand{\ee}{\end{equation}}
\newcommand\beq{\begin{eqnarray}}
\newcommand\eeq{\end{eqnarray}} 
\newcommand\beqs{\begin{eqnarray}}
\newcommand\eeqs{\end{eqnarray}} 
\newcommand\eqn[1]{\label{eq:#1}} 
\newcommand\eq[1]{Eq.~(\ref{eq:#1})}

\newcommand{\GeV}{{\rm ~GeV }}

\newcommand{\eps}{\epsilon}
\def\mag[#1]{\left| #1 \right|}
\def\exv[#1]{\langle #1 \rangle}
\def\mup{{\mu^\prime}}

\begin{document}

\preprint{ITP}

\title{Protecting unparticles from the MSSM Higgs sector}

\author{Ann E. Nelson$^{1,2}$}
\email{anelson@phys.washington.edu}
\author{Maurizio Piai$^{1,3}$}
\email{m.piai@swansea.ac.uk}
\author{Christopher Spitzer$^{1,2}$}
\email{cspitzer@u.washington.edu}

\affiliation{$^1$Dept. of Physics, University of Washington,  Seattle, WA 98195-1560,USA}
\affiliation{
           $^2$Instituto de Fisica Teorica, IFT-UAM/CSIC,  
           Facultad de Ciencias, C-XVI,
           Universidad Aut\'onoma de Madrid,
           Cantoblanco, Madrid 28049, SPAIN}
\affiliation{$^3$Swansea University, School of Physical Sciences,
Singleton Park, Swansea, Wales, UK}

\begin{abstract}
 We construct a model of an unparticle  sector consisting of  a supersymmetric SU(N) gauge theory with the number of flavors in the Seiberg conformal window. We couple this sector to the MSSM via heavy messengers. The resulting low energy theory has a Higgs  coupling to unparticles. The Higgs vev drives the hidden Seiberg sector to a new conformal fixed point. The coupling to the Higgs   mediates supersymmetry breaking to the Seiberg sector, and  breaks conformal invariance  at a lower scale.   The low energy theory contains light stable and metastable mesons. Higgs decay into this sector gives   signatures which are similar to those of ``hidden valley'' models. Decays of the lightest superpartner of standard model particles into the hidden sector reveal potentially observable  unparticle kinematics.
 
\end{abstract}
\date{\today}
\maketitle
\section{Introduction}

There is  a variety of compelling, albeit indirect, evidence for new physics  beyond the minimal standard model.  Such physics might conceivably remain undiscovered, even if produced in high energy collisions,  as finding new physics generally requires understanding what  experimental signatures to look for. There is a standard set of searches which have been developed for  the traditional theories of physics beyond the standard model, such as weakly coupled supersymmetry. It is however important to consider   new physics models with experimental signatures which are as diverse as possible.  Strassler and Zurek  have emphasized the novel signals of ``hidden valleys"---new light sectors with strong self-interactions   which are   weakly coupled to standard model particles at low energies \cite{Strassler:2006im,Strassler:2006ri}. Similarly, Georgi has explored the unusual phenomenological  signals of ``unparticles"---a hidden strongly coupled scale invariant sector which decouples from the standard model at low energy \cite{Georgi:2007ek,Georgi:2007si}.  Although there is no   theoretical necessity for these particular sectors, consideration of such models has the virtue of stimulating new analysis techniques and experimental searches, which are relevant for a broad class of theories.  

Elucidating the observational  features of new strongly coupled sectors is theoretically challenging. One approach, called ``deconstruction" \cite{Stephanov:2007ry}, replaces a strongly coupled scale invariant sector with a closely spaced discrete spectrum of weakly coupled scalars. Another approach is to use  a 5 dimensional dual description \cite{Cacciapaglia:2008ns}. Strassler has criticized these approaches as applying to only an extreme large $N$ limit which may fail  to capture all the phenomenologically important effects \cite{Strassler:2008bv}. Sannino and Zwicky \cite{Sannino:2008nv} considered a walking technicolor model in which both the  unparticle sector  consisted of a gauge theory with sufficient number of massless flavors so as to be at a Banks-Zaks fixed point\cite{Banks:1981nn}. Another option is to consider strongly coupled supersymmetric theories in the Seiberg \cite{Seiberg:1994pq,Intriligator:1995au} conformal window, since supersymmetry can then be used to gain some theoretical knowledge of the dynamics  in a realistic  strongly coupled model \cite{Fox:2007sy}. Using this approach, Fox et al argued that couplings between the unparticle sector and the Higgs sector would push the unparticle dynamics away from the infrared scale invariant fixed point, and introduce a mass scale of order the weak scale into the unparticle sector. The argument that a Higgs-unparticle coupling will drive the unparticle sector away from the conformal fixed point is rather generic \cite{Delgado:2007dx,Kikuchi:2007qd}. In this paper, however, we will illustrate a different possibility, also using a supersymmetric unparticle sector   with a coupling to the Higgs. In contrast with the approach of Fox et al.,    we arrange the coupling to the Higgs  in such a way that the approximate scale invariance of the unparticle sector survives down to an energy scale $\Lambda$ which is parametrically lower than the weak  scale.  Furthermore, we argue that below the scale $\Lambda$ the theory is driven towards a new superconformal fixed point.  A different approximately scale invariant supersymmetric theory   describes the dynamics between $\Lambda$ and a lower scale $f_s$. At still lower energies, below  $f_s$ the theory is a nonsupersymmetric, confining QCD-like theory of light stable and metastable particles. We briefly discuss the possibilities for discovery of unparticles via the decay of the Higgs or   the decay of the lightest superpartner of the Minimal Supersymmetric Standard Model, and the effects of unparticles on Higgs discovery. We also briefly sketch  some of the consequences of   a dark matter candidate in the unparticle sector. We introduce this realistic, predictive model as a laboratory for illustrating and testing generic claims about unparticle   phenomenology.

\section{Model with a  Hidden  Supersymmetric Conformal Sector}

We wish to use a strongly coupled  field theory whose dynamics are well understood for our conformal hidden sector. As a simple, familiar example, we will 
 assume a hidden $SU(N_c)$ gauge theory with $N_f$ flavors of massless quarks.
A variety of consistency checks \cite{Seiberg:1994pq} lead to the conclusion that supersymmetric   $SU(N_c)$  gauge theory in the ``conformal window" with   $N_f$ satisfying 
\be
3/2<\frac{N_f}{N_c}<3  ,
\ee
 must run  to a strongly coupled conformal infrared fixed point in the infrared. Furthermore, this fixed point has dual descriptions--the original ``electric" description, which   is weakly coupled for 
\be\frac{3N_c-  N_f}{N_c}\ll1 ,\ee
  and a ``magnetic" description, with gauge group $SU(N_f -N_c)$ and with the dual quarks coupled to a gauge singlet ``meson"  field. The magnetic description is weakly coupled for 
  \be\frac{N_f-\frac32N_c}{N_c}\ll1 .\ee 
  
  We extend the  Standard Model   to the  Minimal Supersymmetric Standard Model (MSSM) with gauge mediated soft supersymmetry breaking \cite{Dine:1993yw,Dine:1994vc,Dine:1995ag} terms of order the weak scale. To couple the MSSM to the unparticles we introduce heavy gauge singlet superfields $S$ and $T$.
  
 The   superpotential of the model in the magnetic description is
 \beq{\cal W}&=&W_{MSSM} +\lambda \sum_{ij}\bar{q}_i A_{ij} q_j+ \lambda_S S H_u H_d + \mu' S T \nonumber\\&&+ \sum_{ij} g_{ij} {\mu'}^{1-\gamma}S A_{ij} \ . \eeq
 where $W_{MSSM}$ is the superpotential of the MSSM, $H_{u,d}$ are the up and down type Higgs superfields, $q_j,\bar{q}_i $ are the unparticle   quark and anti-quark superfields, $A_{ij}$ is the unparticle meson field, the index $i$ runs from 1 to $N_f$,  and  we have suppressed the color indices. 
We assume the number of unparticle quark flavors is in the conformal window.
 In the absence of the deformations induced by the visible sector, 
there is an infrared attractive fixed point in the hidden sector, where the theory is described by
 the conformal theory based on operators with scaling dimension
  $d(A_{ij})= 1+\gamma$, 
 and $d(\bar{q}_iq_j)=2-\gamma$, with  $\gamma $ constant. We assume that   $\mu'$ is well above the weak scale, of order a few TeV, and that at this high scale the unparticle sector is  nearly at the conformal  fixed point. 
 In our  analysis we will assume that the matrix $g_{ij}$ has
non-vanishing entries only for $i,j=N_f-n+1,\cdots,N_f$, 
all of the same order of magnitude (but different). Without loss of generality, we may use flavor transformations to diagonalize $g_{ij},$, and make the diagonal entries real and positive. The factor of ${\mu'}^{1-\gamma}$ is inserted so that
 the $g_{ij},$ are effectively dimensionless. Here the couplings $g_{ij}$ are renormalized   at the scale $\mu'  $ , and are assumed to be perturbative at this scale. We may also take the weak coupling $\lambda_S$ to be real and positive.

 The approximate  global flavor symmetry of  the unparticle sector is explicitly broken by the $g_{ii}$ couplings to  a global $SU(N_f-n)\times SU(N_f-n) \times U(1)^n\times U(1)_B\times U(1)_R$.    (The $U(1)_R$ symmetry is not exact when supersymmetry breaking effects are considered.)
 
 \subsection{Effective Field Theory Analysis}
 
 We now discuss the low energy effective theory below the scale $\mu'$, at which we integrate out the $S$ and $T$ superfields. 
 We assume that $\mu^{\prime}$ is of order a TeV, large compared with the weak scale and the MSSM  susy breaking scale. We also assume that the mediation of susy breaking to the hidden sector occurs indirectly via the $S$ and $T$ couplings, and so the scale of susy breaking in the hidden sector is suppressed below the weak scale.    We can therefore   neglect supersymmetry breaking as we  integrate them out. 
Note that the $S$ and $T$ fields only have weak couplings, and so it is a good approximation to integrate them out at tree level, by solving the  
 classical equations which result from minimizing the potential.    The equations of motion are 
\beq
F_S&=&F_T\,=\,0\,,\\
S&=&0\,,\\
T&=&-\frac{1}{\mu^{\prime}}\left(\lambda_SH_uH_d+\sum_ig_{ii}{\mu'}^{1-\gamma}A_{ii}\right)\,.
\eeq
 The resulting effective theory has a superpotential
\beq
{\cal W}&=&W_{MSSM} + \lambda\bar{q}_i A_{ij} q_j \,. 
\eeq 
Unlike in the model of Fox et al., there is now no direct superpotential coupling between the hidden and visible sectors.   Note that this lack of direct coupling results from our inclusion of the $T$ superfield with a  nongeneric  superpotential. The specific form of our superpotential is technically natural. 

The low energy effective theory does couple the hidden and MSSM sectors, but only via terms in the effective Kahler potential. These terms induce an effective supersymmetric superpotential term in the hidden sector Lagrangian, in a manner reminiscent of the Giudice-Masiero \cite{Giudice:1988yz} mechanism for generating the $\mu$ term in the MSSM.
The additional terms arising
at the leading order in an expansion in powers of $1/\mu^{\prime}$ in the Kahler potential are
obtained by replacing the solutions for $T$ and $S$:
\beq
{\cal K}_{\rm eff}&\supset& S^{\dagger}S\,+\,T^{\dagger}T\,
\nonumber\\
&=&\frac{1}{\mu^{\prime\,2}} \left|\lambda_SH_uH_d+\sum_ig_{ii}{\mu'}^{1-\gamma}A_{ii}\right|^{2}
\, \nonumber\\
\label{effkahler}&=&
\left|\sum_i\frac{g_{ii}}{\mu^{\prime\,\gamma}} A_{ii}\right|^2
\,+\,\frac{\lambda_S^2}{\mu^{\prime\,2}}\left| H_uH_d\right|^2\,\\
&&\,+\,\sum_i\frac{\lambda_Sg_{ii}}{\mu^{\prime\,1+\gamma}}\left(H_uH_dA^\dagger_{ii}+H_u^\dagger H_d^\dagger A_{ii}\right)\ .\nonumber\eeq
We now run this effective theory down to the weak scale $v\equiv175$ GeV, which we take to also be of order the scale of susy breaking in the MSSM sector (the effective susy breaking scale in the hidden sector will be much smaller). 

The superpotential terms do not run, since the MSSM terms are weakly coupled, while the hidden sector terms are at a fixed point. We do not know how the   first term in Eq. (\ref{effkahler}) runs, but we do know it is an irrelevant term with negligible effect on hidden sector dynamics at all scales below $\mu'$. The second term only involves weakly coupled fields and does not run. The third term in Eq. (\ref{effkahler}), which in linear in the strongly coupled fields $A_{ij}$, has anomalous dimension $\gamma$. At the scale $v$ this term becomes
\be\label{effkahler2}{\cal K}_{\rm eff}\supset \frac{v^\gamma}{\mu^{\prime\,1+2 \gamma}}\left(\lambda_Sg_{ii}H_uH_dA^\dagger_{ii}+\lambda_Sg_{ii}H_u^\dagger H_d^\dagger A_{ii}\right)\ . \ee
At the scale $v$ we integrate out the Higgs superfields. These are weakly coupled fields, and we can integrate them out at tree level by setting the scalar and $F$ components to their vacuum values.  We define the parameters
\be\epsilon_{ii}\equiv  \lambda_Sg_{ii} \left(\frac{v}{\mu'}\right)^{1+2\gamma},\eqn{epsdef}\ee  which control the strength of  quantum corrections in the hidden sector due to the coupling between the two sectors, and take $\epsilon_{ii}\ll1$. Quantum loop corrections    are suppressed by    the $\epsilon$ coefficient squared, times a loop  factor, in units of the scale $v$. To leading order in the $\epsilon$ parameters, the Higgs scalar and $F$ term expectation values are independent of the hidden sector fields. Neglecting tree level and quantum contributions of order $\epsilon^2$,  the only terms  from the MSSM sector that survive in Eq. (\ref{effkahler2})  are proportional to  $\langle F_{H_uH_d}\rangle $.  (Because of the soft weak scale supersymmetry breaking in the MSSM sector, we do not set the  $F$ terms  for the Higgs fields to zero. )  
Below the weak scale, we can  then rewrite the   term (\ref{effkahler2}) as a contribution to the effective superpotential for the hidden sector
 
\beq\label{effsuper}
{\cal W}_{\rm eff}&=& \lambda\sum_{ij}\bar{q}_i A_{ij} q_j + \sum_i\frac{\langle F_{H_uH_d}\rangle }{v^{1+\gamma}} \epsilon_{ii} A_{ii} \ ,\eeq
where
\beqs
\langle H_u \rangle &=& v_W \sin\beta\,,\\
\langle H_d \rangle &=& v_W \cos\beta\,,\\
\langle F_{H_u} \rangle &=&  \mu_W v_W\cos\beta \,,\\
\langle F_{H_d} \rangle &=&  \mu_W v_W\sin\beta \, \\
\langle F_{H_uH_d}\rangle&=& \langle H_u \rangle \langle F_{H_d} \rangle+\langle H_d \rangle \langle F_{H_u}\rangle\ ,\eeqs
and $\mu_W$ is the usual $\mu$ term of the MSSM. 
Remarkably,  to leading order in the coupling between the hidden and visible sectors, the supersymmetry breaking in the MSSM leads to a term in the effective theory for the hidden sector which can be written as   supersymmetric, although not scale invariant. Because this term is   relevant, it will drive the hidden sector away from the fixed point in the infrared. However, because the theory is still effectively supersymmetric, we still have theoretical control over the dynamics. In a supersymmetric theory with a supersymmetric vacuum, the vacuum may be found by setting the  $F $ terms derived from the effective superpotential to zero, along D-flat directions. 
\beq\label{vacuum}
\langle \bar{q}_{i}  {q}_{i} \rangle =  {\epsilon_{ii}\frac{\langle F_{H_uH_d}\rangle }{\lambda v^{1+\gamma}} }\,,
\eeq 
In this vacuum, $n$ flavors of magnetic quarks obtain expectation values, partially higgsing the magnetic gauge group.
In Eq. (\ref{vacuum}), all fields and couplings are renormalized at the scale $v$. As long as $n$ is not too large,  that is, as long as  
\be \frac32<\frac{N_f-n}{N_c}<3\ ,\ee the theory flows to a new fixed point, with different but still nontrivial anomalous dimensions.
To find the physical scale $\Lambda$ at which the theory is driven to a new fixed point, we note  between $v$ and $\Lambda$, the   $A $  fields have anomalous dimension $ \gamma$. We   take $F_{H_uH_d}$ to be of order $v^3$, and  all the $\epsilon_{ii}$ to be of  similar size.  We can  determine  the order of magnitude of $\Lambda $  by using dimensional analysis. At the scale $\Lambda$ the  renormalized coefficient of $A_{ii}$  in Eq. (\ref{effsuper} ) is of order $\Lambda^{2-\gamma}$. We can therefore use   \be \epsilon_{ii}   v^{2-\gamma}\sim\Lambda^{2-\gamma}\ee to get
\be\Lambda\sim \epsilon_{ii}^{\frac{1}{2-\gamma}} v\ .\ee Note that as long as the $\epsilon_{ii}$ parameters are less than 1, $\Lambda$ is below the weak scale. Furthermore, the more strongly coupled the magnetic description is (the closer $\gamma$ is to 1) the smaller $\Lambda$ is.
Loop effects and tree effects which are higher order  in the weak $\epsilon_{ii}$ couplings will induce supersymmetry breaking effects into the hidden sector,   at a   scale $f_s<\Lambda$.  At the weak scale, such non supersymmetric tree  corrections are proportional to  a factor of $\epsilon_{ii}^2$, times powers of $v$ given by dimensional analysis, and loop corrections are roughly proportional to powers of $\epsilon_{ii}^2/(16 \pi)^2$. A rough estimate for the upper bound on the scale  of susy breaking in the hidden sector   is   $ (\epsilon    v)$,  where $\epsilon= $Max$\epsilon_{ii}$, which is  lower than the scale  $\Lambda$   by a factor of $ {\epsilon}^{\frac{1-\gamma}{2-\gamma}}$. 

We have several theoretical handles on the  effective theory between $f_s$ and  $\Lambda$. In the limit that $N_f-n$ of the $g_{ii}$ couplings vanish, this theory possesses  an $SU(N_f-n)\times SU(N_f-n)$ flavor symmetry, which is also respected by the  supersymmetry breaking terms for the fields in this effective theory. In reference \cite{Nelson:2001mq}, Nelson and Strassler showed that all soft   supersymmetry breaking terms in   a superconformal effective theory which respect the  flavor symmetry have positive anomalous dimension.  Certain linear combinations of quark masses  squared, i.e. those proportional to global symmetry generators, do not run \cite{Nelson:2001mq}, however  generating such symmetry violating terms is not possible when the supersymmetry breaking is mediated by flavor symmetric couplings.  The anomalous dimensions for the supersymmetry breaking terms cannot be nonperturbatively computed, but, since these are known to be positive at an  infrared attractive  fixed point, renormalization group scaling will lower the supersymmetry breaking scale $f_s$ below $ (\epsilon   v)$.  Furthermore if the infrared fixed point is weakly coupled in the magnetic description, as is possible at large $N_c$ with the appropriate number of flavors, the scale of susy breaking will be suppressed by additional factors of $1/N_c$. (In this limit, however, the anomalous dimension $\gamma$ is   small, making   unparticle kinematics less exotic.) We conclude that  in the small $\epsilon$ limit the physical supersymmetry breaking scale $f_s$ is significantly below the scale $\Lambda$. We treat $f_s$ as a free parameter subject to the $(\epsilon   v)$ upper bound, although in principle it is determined by the parameters and the strong dynamics.  Below $f_s$,   we assume the unparticle sector is no longer conformal, and, depending on the sign of the soft supersymmetry breaking squark masses,   either the magnetic description is in a Higgs phase  and the electric description in   a QCD-like confining phase, or vice-versa. 

\subsection{Hidden Sector Spectrum}

As seen in the previous section, electroweak symmetry breaking introduces several scales into the hidden sector, and  a spectrum of strongly coupled particles. Most of these are unstable, however some will be stable due to unbroken flavor symmetries.   In this section we give an overview  of the long lived and stable spectrum of particle states.

We begin with the lightest particles.
Below the scale $f_s$ we have a strongly coupled nonsupersymmetric theory, with a global $SU(N_f- n)\times SU(N_f- n) \times U(1)_B $ flavor symmetry, where $U(1)_B$ refers to baryon number. We assume that either the magnetic or the electric description of the low energy nonsupersymmetric effective theory  is  confining at the scale $f_s$ . The global symmetries of the theory are a chiral $SU(N_f-n)\times SU(N_f-n)$, and  a $U(1)_B$ baryon number. Depending on the value of the gluino mass, there may also be an approximate spontaneously broken chiral $U(1)$ symmetry, with the anomaly from the quark sector cancelled by the gluino.  
We will assume QCD-like dynamics, with spontaneous breaking of the chiral  $SU(N_f-n)\times SU(N_f-n)$  symmetry to the diagonal subgroup, giving $(N_f- n)^2-1$ massless Nambu-Goldstone bosons.   

The chiral symmetry  is due to our assumption that the coupling matrix $g_{ij}$ is rank $n$. The discussion of the previous section is still valid when the matrix $g$ is rank $N_f$, provided  there is a hierarchy in the couplings, with $n$ of the eigenvalues are of the same order of magnitude and with the other eigenvalues much smaller, so they don't affect the dynamics at the scales $\Lambda$ or $f_s$.  Such small couplings explicitly break  $SU(N_f- n)\times SU(N_f- n) \times U(1)_B $ to the diagonal $U(1)^{N_f- n} $, and give small masses to all the Nambu-Goldstone bosons, of order $\sqrt{g_{jj}}\xi f_s$, where the $g_{jj} $ are the smaller couplings which were neglected in the previous section, and $\xi$ is a factor from scaling the $g_{jj}$ couplings.    The lightest particles of the theory are therefore a multiplet of light pseudoscalar   bosons, analogous to the pions, kaons and eta of QCD. The unbroken $U(1)$ symmetries will make $(N_f-n-1)(N_f-n) $ of these stable, while the remaining  $N_f-n-1$  pseudoscalars can mix with the pseudoscalar Higgs and decay to lighter standard model particles. Other stable particles of the theory are the baryons, with masses of order $ N_c f_s$.

We also expect heavier particles associated with the scale $\Lambda$, the scale associated with the
  condensate of some of the  magnetic quarks,
which partially breaks both the global and the magnetic-color symmetries. These particles are analogous to the mesons   of QCD containing heavy quarks, with the novel feature that they come in approximately degenerate supermultiplets.  Note that an exact diagonal $U(1)^n$ however  unbroken, yielding additional stable  and mesons and baryons with mass of order $\Lambda$. There is also    a spectrum of unstable resonances.  

We can  be more definite about the spectrum in the limit where the magnetic description is weakly coupled (see also the Appendix for a discussion of large $N_c$ scaling). In this weakly coupled limit,  at the scale $\Lambda$ the magnetic $SU(N_c-N_f)$ gauge symmetry is Higgsed down to $SU(N_c-N_f-n)$, and the massive particles at the scale $\Lambda$ come in   vector supermultiplets. There are $n^2$ vector supermultiplets which are uncharged under the low energy gauge group, and $n$ vector supermultiplets transforming in the fundamental and anti fundamental representations, which at long distances will form bound states with each other and with lighter particles. At the scale $f_s$, small splittings are introduced into the  spectrum of heavy supermultiplets. 

In a previous paper \cite{Nelson:2008hj}, two of us considered a dark matter sector containing heavy stable   and lighter unstable particles, and discussed how the PAMELA \cite{Adriani:2008zr} cosmic ray positron spectrum could result from annihilation of the heavy stable dark matter into the lighter particles, which then decay into electron positron pairs. This model exhibits all the necessary ingredients of such a dark matter model, if the heavy stable particles are identified as dark matter, and the scale $\Lambda$ is above 80 GeV. In addition, the dark matter states come in multiplets with small splittings, which may allow for the inelastic dark matter explanation \cite{TuckerSmith:2004jv} of the DAMA/LIBRA experiment \cite{Bernabei:2008yi} and/or the exciting dark matter \cite{Finkbeiner:2007kk} explanation of the Integral low energy positron excess \cite{Knodlseder:2005yq}.

One final comment concerns the LSP, or lightest superpartner. As in all gauge mediated models, the LSP will be an ultralight gravitino, and the next lightest superpartner, or NLSP, will decay into the gravitino. Typically in gauge mediated models the NLSP is one of the MSSM states, e.g. the stau or the lightest neutralino. In this model, the NLSP will be a fermion in the hidden sector, e.g. a bound state of a scalar and a fermion with mass of order $f_s$, times small couplings. The lightest MSSM state will therefore decay into the hidden sector, implying that every collider event that produces superpartners will produce hidden sector unparticles. 
The NLSP in turn will decay into a gravitino and one or more pseudoscalar mesons. Some of the  latter particles will decay back into the visible sector via mixing with the pseudoscalar Higgs, making some NLSP decays potentially visible.

\subsection{Comparing to NDA.}

We summarize here some of the arguments in~\cite{Fox:2007sy, Bander:2007nd},
in order to clarify the differences with the set-up of this paper.
In a generic realization of the unparticle scenario, the couplings between 
the standard model and the unparticle sector can be described in the far UV
by a set of higher-order operators:
\beqs
c_n^i\frac{O_{UV}O_n^i}{M^{d_{UV}+n-4}}\,,
\label{Eq:UVcoupling}
\eeqs
where we used the same notation as in~\cite{Bander:2007nd}, according to which
$O_{UV}$ is
an operator of dimension $d_{UV}$ in the hidden sector, $O_n^i$ is a
standard-model operator of dimension $n$, $c_n^i$ is a dimensionless
coupling and $M$ some very high scale.
The latter may be thought of as the mass of the fields responsible for
mediating the interactions between the two sectors (in our set-up, this
would be related to the mass scale $\mu^{\prime}$).

Below the scale $\Lambda_{U}$ at which the hidden sector enters its
conformal phase,
the operator $O_{UV}$ flows onto a $d$-dimensional operator $O$, resulting
in the 
effective coupling
\beqs
c_n^i\frac{O O_N^i}{\Lambda_n^{d+n-4}}\,,
\label{Eq:IRcoupling}
\eeqs
where $\Lambda_n^{d+n-4}\equiv M^{d_{UV}+n-4}/\Lambda_U^{d_{UV}-d}$.
ª
The problem highlighted in~\cite{Bander:2007nd} is that one of the operators 
of the standard model is $O_2=H^{\dagger}H$, the $n=2$ dimensional bilinear 
in the Higgs sector. Also, the operators $O$ that are most interesting
phenomenologically, 
and most often considered in the literature, are
unparticle operators with $1<d<2$. This results in a relevant effective
interaction 
from Eq.~(\ref{Eq:IRcoupling}). Furthermore, 
below the electro-weak scale, it induces the breaking of conformal symmetry
in the
unparticle sector by a highly-relevant operator.

The conclusion is that  a generic unparticle model
cannot yield sizable phenomenological signals
unless fine-tuning is introduced in order to protect the unparticle sector
from the
effect of the scale of electro-weak symmetry breaking.
Only at large scales can the unparticle description hold,
while at scales parametrically lower than the electro-weak scale the
unparticle sector 
turns into a hidden sector of massive, neutral particles with conventional
phenomenology.

In this paper we took a closer look at the mechanism generating the
couplings in
Eq.~(\ref{Eq:UVcoupling}), in order to gauge the generality of the 
conclusion
summarized above. 
Instead of allowing for generic unconstrained couplings,
we explicitly built a model both for the unparticle sector (based on a
${\cal N}=1$ $SU(N_c)$
gauge theory) and for the mediators (the $S$ and $T$ superfields), which,
in order to preserve supersymmetry and hence allow for calculability, 
we coupled to the MSSM in the visible sector, with a very specific choice of
superpotential ${\cal W}$.
Soft terms in the MSSM are the only explicit source of susy breaking.

In comparison with~\cite{Bander:2007nd}, we find three important differences.
\begin{itemize}
\item  Supersymmetry, and hence a degree of calculability, is preserved in
the
hidden sector down to a scale parametrically lower than the electro-weak
scale.
\item The scale induced in the hidden sector by the Higgs VEV  in the
visible sector
is itself parametrically suppressed in respect to the electro-weak scale.
\item Below the latter scale, conformal symmetry is broken only in the sense
that 
the hidden sector flows away from the IR fixed point it is approaching below
$\Lambda_U$,
but is recovered asymptotically as in the far IR the RG flow approaches a
different fixed point.
\end{itemize}

Some important  physical implications of these three results.
First of all, the analysis we carried out  is reliable enough that 
we do not need to introduce generic NDA arguments.
A large energy range over which the hidden sector behaves as a genuine
unparticle sector 
exists, down to energies well below the electroweak scale.
At even lower energies, not all of the hidden sector degrees of freedom 
become just ordinary 
 massive particles, but a new conformal regime appears.

We draw three final lessons.
First of all, the analysis of~\cite{Bander:2007nd} is correct for generic models, but
being only based on NDA arguments it cannot be applied
blindly to all possible models: our work shows that there are specific 
realizations of the mediation mechanism that evade the bounds in~\cite{Bander:2007nd}
without fine-tuning.

On the other hand, the specific example we constructed is very peculiar: 
besides being supersymmetric (which in our context is only a technical
requirement),
it requires the interplay of two singlet fields $S$ and $T$, 
with a very specific $W$,
and cannot work in presence of only one of these mediators. 
In other words: it is not easy to construct counterexamples 
to the pessimistic conclusions of~\cite{Bander:2007nd}.
The good implication is that if experimental data were to discover
unparticle-type 
signatures, these would yield interesting information not only about the
unparticle sector itself,
but also about the mechanism that couples it to the visible sector at high
energy.

Finally, the third lesson is that at low energies the phenomenology of these
models
is very rich, consisting of an admixture of particle- and unparticle-like
signatures,
where part of the new particle content is determined by the transition between two
different IR
fixed points induced indirectly by the electro-weak symmetry breaking VEV.
\section{Observing Unparticles}

\subsection{Higgs-to-unparticle couplings}

When the theory described in previous sections is at its fixed point, the quark and meson fields of the hidden sector attain fixed anomalous dimensions, and we obtain couplings to fields recently described as unparticles.  At tree level, the only coupling of the MSSM to this new physics is between the Higgs and unparticle operators.  In this section we will consider the leading order couplings to neutral Higgs.  We will take some number $n$ of nonzero couplings, that is
\be
g_{ii} \not= 0
\ee
where $(N_f-n+1) \leq i \leq N_f$, and we take $g_{ii}$ to be perturbative.

We now determine how unparticles couple to scalar Higgs mass eigenstates.  In the MSSM, the mixing of the Higgs fields $H_u$ and $H_d$ into these fields can be described by
\be
H_u=\left(\frac{\sqrt{2}v_u+R_u}{2}\right)\exp\left\{i\frac{I_u}{\sqrt{2}v_u}\right\} \eqn{huexp},
\ee
\be
H_d=\left(\frac{\sqrt{2}v_d+R_d}{2}\right)\exp\left\{i\frac{I_d}{\sqrt{2}v_d}\right\} \eqn{hdexp}
\ee
where
\be
R_u=\cos\alpha h^0+\sin\alpha H^0,
\ee
\be
R_d=-\sin\alpha h^0+\cos\alpha H^0,
\ee
\be
I_u=\cos\beta A^0,
\ee
\be
I_d=\sin\beta A^0,
\ee
where we have gone to unitary gauge.
The fields $h^0$ and $H^0$ are the light and heavy Higgs scalars, and $A^0$ is the pseudoscalar Higgs.  To leading order the mixing angles $\alpha$ and $\beta$ are determined by the parameters of the Higgs sector, and match the MSSM values in absence of the hidden sector.  The terms $v_u$ and $v_d$ are the expectations values for $H_u$ and $H_d$, and are also equivalent to the typical MSSM values up to small corrections.

We will work with the effective potential which has only the heavy scalars $S$ and $T$ integrated out, described in previous sections as $\mathcal{K}_{\mathrm{eff}}$ and $\mathcal{W}_{\mathrm{eff}}$.  We will make explicit the presence of quantum corrections in the Kahler potential, and include effects of the running of $g_{ij}$ to scale $\mu$.  This leaves us with an effective Kahler potential of the form
\beq
\mathcal{K} & \supset & \frac{\lambda_S^2}{\mup^2}\mag[H_uH_d]^2 +\frac{\lambda_S \mu^\gamma}{\mup^{1+2\gamma}}(H_uH_d\sum_ig_{ii}A_{ii}^\dagger +\mathrm{h.c.}) \nonumber \\
 & & + K(A_{ij},q_i,\bar{q}_i,A_{ij}^\dagger,q_i^\dagger,\bar{q}_i^\dagger).
\eeq
Here $K$ is the Kahler potential of the hidden sector, including quantum corrections which may, in general, be a complex function of the fields.  Subscripts on $K$ refer to derivatives with respect to the field given in the subscript.

Upon integration of the superspace variables, the cross-terms in the full potential $\mathcal{K}$ generate a number of Higgs-unparticle couplings in the Lagrangian.  These have the schematic form
\be
\mathcal{L} \supset \mathcal{L}_{Hq} + \mathcal{L}_{HA}
\ee
where the first term contains couplings to the hidden quark fields, and the second to the hidden meson fields.

The coupling between Higgs and quarks is generated by
\beq
\mathcal{L}_{Hq} &=& \sum_i \frac{g_{ii}\lambda_s\mu^\gamma}{\mup^{1+2\gamma}} F_{A_{ii}A_{ii}}(H_u^* F_{H_d}^*+H_d^* F_{H_u}^*+2\Psi_{H_u}^\dagger \Psi_{H_d}^\dagger) \nonumber \\
& & +\mathrm{h.c.}
\eeq
where the $H$ fields are now scalars, and the $\Psi$ fields are fermionic.  Solving the equation of motion for the $A_{ii}$ auxiliary field yields the solution
\be
F_{A_{ii}} = -\sum_j  K_{\Phi^*_jA_{ii}}^{-1} W_{\Phi_j}^* \equiv \mathcal{O}_i
\ee
where the sum runs over all fields $\Phi^*_j$ on which $K$ has functional dependence.  The dimension of this unparticle operator, $\mathcal{O}_i$, matches the dimension of the $A_{ii}$ field F-term, which is $2+\gamma$.
Inserting these solutions in to the Lagrangian yields the following couplings,
\beq
\mathcal{L}_{Hq} & =  & \sum_i \frac{g_{ii} \lambda_s \mu_W\mu^\gamma}{\mup^{1+2\gamma}}(\mag[H_u]^2+\mag[H_d]^2)\left( \mathcal{O}_i + \mathrm{h.c.} \right) \nonumber \\
 & & + \sum_i \frac{2g_{ii} \lambda_s \mu^\gamma}{\mup^{1+2\gamma}}\Psi_{H_u}^\dagger \Psi_{H_d}^\dagger \mathcal{O}_i + \mathrm{h.c.}.
\eeq
Expanding this to Higgs mass states at leading order provides the Higgs-unparticle couplings,
\beq
\mathcal{L}_{Hq} &= & \frac{ \lambda_s \mu_W \mu^\gamma}{4 \mup^{1+2\gamma}} \Big\{ {h^0}^2 + {H^0}^2 + 2\sqrt{2}{H^0}v \cos(\alpha-\beta) \nonumber \\
&& - 2\sqrt{2}{h^0}v\sin(\alpha-\beta) \Big\} \sum_i (g_{ii}\mathcal{O}_i+\mathrm{h.c.}) \nonumber \\
&& + \sum_i \frac{2 g_{ii}\lambda_s\mu^\gamma}{\mup^{1+2\gamma}} \Psi_{H_u}^\dagger \Psi_{H_d}^\dagger \mathcal{O}_i + \mathrm{h.c.} \label{eq:explang}
\eeq
where $v^2=v_u^2+v_d^2$.

There are also interactions between Higgs and the hidden meson fields that come from the $A$ cross-terms in the Kahler potential.  Integrating over superspace yields 3-scalar terms and fermion-fermion-scalar terms
\beq
\mathcal{L}_{HA}&\supset& \sum_i \frac{\lambda_s g_{ii} \mu^\gamma}{\mup^{1+2\gamma}} \Big\{ \nonumber \\
 & &-\left( \frac{3}{2}H_u^*H_d^*\partial^2 A_{ii} + \frac{1}{2} A_{ii}\sigma^\mu\partial_\mu H_d^* \sigma^\nu \partial_\nu H_u^* \right) \nonumber \\
 & & +\left( 2i\sigma^\mu\partial_\mu H_u^* \Psi_{H_d}^\dagger \Psi_{A_{ii}}  + 2i\sigma^\mu \partial_\mu H_d^* \Psi_{H_u}^\dagger \Psi_{A_{ii}} \right) \nonumber \\
 & & +\left( i H_u^* \Psi_{H_d}^\dagger\partial_\mu\Psi_{A_{ii}}\sigma^\mu + i H_d^* \Psi_{H_u}^\dagger\partial_\mu\Psi_{A_{ii}}\sigma^\mu \right) \nonumber \\
 & & -\left( H_u^* \sigma^\mu\partial_\mu \Psi_{H_d}^\dagger \Psi_{A_{ii}} + H_d^* \sigma^\mu\partial_\mu \Psi_{H_u}^\dagger \Psi_{A_{ii}} \right) \nonumber \\
 & & \Big\} + \mathrm{h.c.} \label{eq:lha1}
\eeq
after combining some terms using integration by parts.

\subsection{Decays to unparticles}

We now briefly examine a few of the processes that are generated by the couplings described above.   Our goal is to point out a few examples of processes with unparticle-like properties, rather than exhaust the space of interesting decay topologies.   These processes occur when the Higgs couples to effectively conformal fields, that is at energies above $\Lambda$, as discussed in previous sections.   As before, we focus on the neutral Higgs scalars which have interesting phenomenological potential at the LHC.  For simplicity of presentation, in this section we will set $n=1$, and define $\mathcal{O}\equiv \mathcal{O}_{N_f}$, $g \equiv g_{NfNf}$, $\eps \equiv \eps_{NfNf}$ and $A \equiv A_{NfNf}$.

The calculations of rates involving unparticles differs somewhat from the standard particle cases due to the unparticle phase space. Following the notation of~\cite{Georgi:2007ek} the phase space for scalar unparticles is
\be
\delta \Phi = A_{d_u}\theta(p_{\mathcal{U}}^0)\theta(p_{\mathcal{U}}^2)(p_{\mathcal{U}}^2)^{d_u-2}
\ee
and for fermionic unparticles (see~\cite{Luo:2007bq,Basu:2008rd,Liao:2008tj}) is
\be
\delta \Phi= \frac{3}{2}A_{2d_u/3}\theta(p_{\mathcal{U}}^0)\theta(p_{\mathcal{U}}^2) (p_{\mathcal{U}}^2)^{d_u-5/2}.
\ee
In each case, $d_u$ is the dimension of an unparticle operator $\mathcal{O}_u$, and $A_{n}$ is a normalization constant given by
\be
A_{n}=\frac{16\pi^\frac{5}{2}\Gamma(n+\frac{1}{2})}{(2\pi)^{2n}\Gamma(n-1)\Gamma(2n)},
\ee
which shouldn't be confused with the meson field $A$.

As a simple example, for a coupling of a scalar and unparticles of the generic form
\be
c h \mathcal{O}_u, 
\ee
where $h$ is scalar with mass $M$ and $c$ is a dimensionful coupling constant, the total decay rate for $H$ into unparticle stuff is given by
\be
\Gamma=\frac{\mag[c]^2}{2}A_{d_u}M^{2d_u-5} \label{eq:scalardecayrate}.
\ee
Note that this approaches the standard decay rate to two massless particles as $d_u$ goes to 2.  If the unparticle operator is derivatively coupled to the scalar, $ch\partial^2\mathcal{O}_u$, the decay rate is
\be
\Gamma=\frac{\mag[c]^2}{2}A_{du}M^{2d_u-1}.
\ee
An alternate method is   to use the optical theorem in a  $1\rightarrow1$ scattering, 
and find the width from the imaginary part of the propagators,
after  including the mixing in  in the unparticle-Higgs system, as suggested in \cite{Delgado:2008gj}. Our procedure is equivalent, and   convenient for the present purposes. 

The first process we consider is the decay of the light Higgs of the MSSM directly into unparticles.  This process is generated by the terms of the form $h^0\mathcal{O}$ in equation (\ref{eq:explang}), and of the form $h^0 \partial^2 A_{ii}$ from terms in equation (\ref{eq:lha1}).  After expanding in terms of mass eigenstates, the relevant coupling terms are
\beq
\mathcal{L} &\supset& -\frac{g \lambda_s \mu_W \mu^\gamma}{\sqrt{2}\mup^{1+2\gamma}}v\sin(\alpha-\beta)h^0\mathcal{O} \nonumber \\
 & & -\frac{3\sqrt{2}}{8}\frac{g\lambda_s\mu^\gamma}{\mup^{1+2\gamma}}v\cos(\alpha+\beta)h^0(\partial^2 A+\mathrm{h.c.}).
\eeq
Using the decay rates from above, we find the total decay rate to unparticles is given by
\beq
\Gamma_{h\rightarrow \mathrm{unp}}=&\frac{g^2\lambda_s^2v^2}{4\mup^{2+4\gamma}} & \big\{  A_{2+\gamma}\mu_W^2\sin^2(\alpha-\beta)m_h^{4\gamma-1} \nonumber \\
&  &+ \frac{9}{16}A_{1+\gamma}\cos^2(\alpha+\beta)m_h^{4\gamma+1} \big\} \nonumber \\
=& \frac{\eps^2}{4}\left(\frac{m_h}{v} \right)^{4\gamma} & \big\{A_{2+\gamma}\mu_W^2\sin^2(\alpha-\beta)m_h^{-1} \nonumber \\
&  &+ \frac{9}{16}A_{1+\gamma}\cos^2(\alpha+\beta)m_h \big\}
\eeq
in which the $\eps$ is now evaluated at the Higgs mass rather than the weak scale, as had been written in \eq{epsdef}.

\begin{figure}[htpb]
\includegraphics[width=8cm]{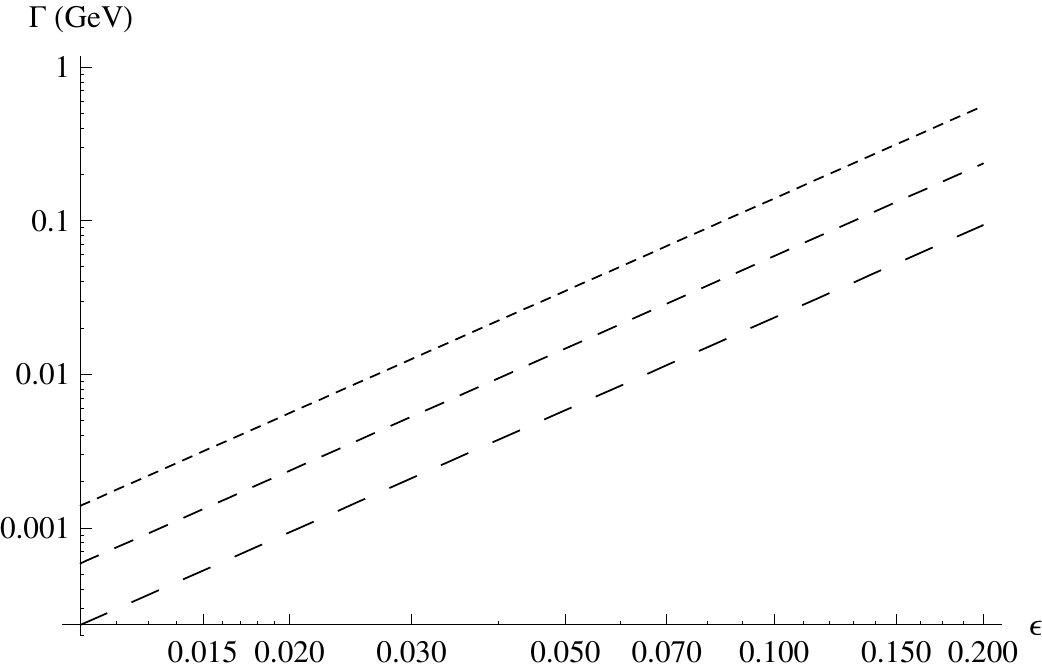}
\caption{The decay rate for the light Higgs to unparticles for the parameters described in the text.  Three values of the anomalous dimension $\gamma$ are shown.  From shortest to longest dashes, these are $\gamma$=0.1, 0.3 and 0.5.  The value of the Higgs mass is fixed at $m_h=120\GeV$.}
\label{fig:pheno0}
\end{figure}

In order to present numerical results, we need to choose a set of MSSM parameters.  We will take $\tan\beta\approx10$ and the decoupling limit $m_{A^0} \gg m_Z$.  Specifically, we set $m_{A^0} \approx 800 \GeV$.  At this point, the tree-level value of the light Higgs mass is $89.3\GeV$, but will receive large corrections at one-loop from couplings to the quarks.  We will take the appropriate squark masses so that the physical value of the light Higgs mass is 120 $\GeV$.  The rate for this decay is show in figure \ref{fig:pheno0}, in which we vary the dimensionless combination of parameters described by $\eps$ at several fixed values of $\gamma$.

At this point in the MSSM parameter space, the couplings of the light Higgs are Standard Model-like, and the dominant decay mode for $h^0$ is to $b\bar{b}$.  The relatively weak coupling of the Higgs and $b$ means that couplings to new sectors do not need to be particularly strong to compete with standard model channels (a concise summary of the situation is presented in~\cite{Chang:2008cw}).  The decay rate for the light Higgs to $b\bar{b}$ at tree level is given by
\be
\Gamma_{h\rightarrow b\bar{b}} = \frac{3}{\pi}\frac{m_b^5}{v^2 m_h^2}\left( \frac{m_h^2}{4m_b^2}-1 \right)^\frac{3}{2}.
\ee
\begin{figure}[htpb]
\includegraphics[width=8cm]{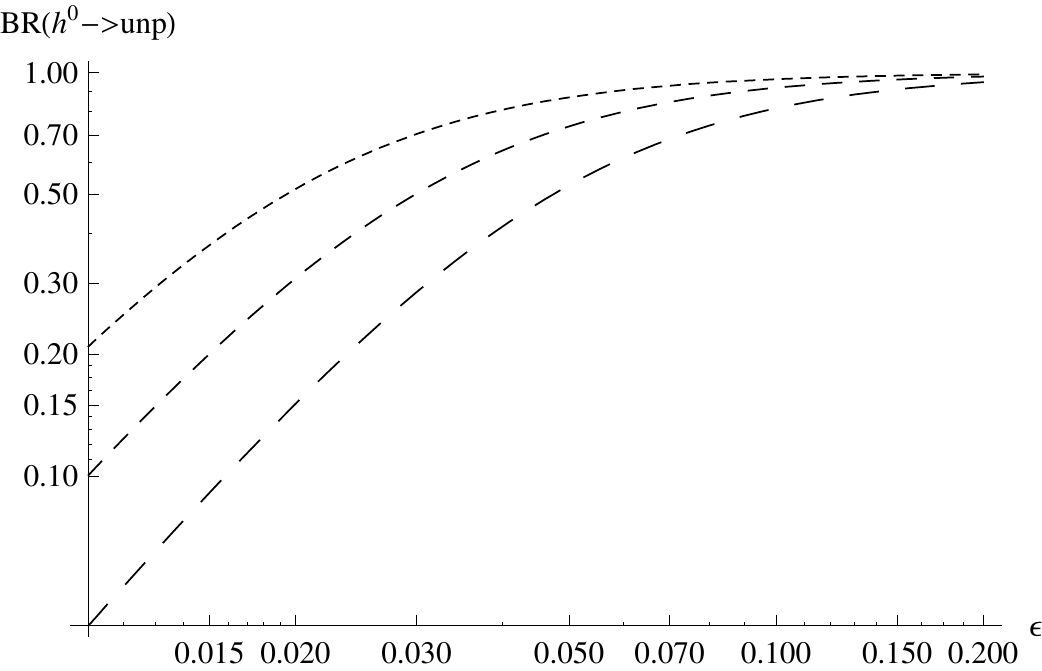}
\caption{The branching ration for Higgs decays to the unparticles.  Three values of $\gamma$ are shown.  From shortest to longest dashing, these are $\gamma$=0.1, 0.3 and 0.5.  The light Higgs mass is fixed at $m_h=120\GeV$.}
\label{fig:pheno1}
\end{figure}
Decays to unparticles will dominate decays to $b$ quarks for certain choices of hidden sector parameters.  We plot this as a bound on the dimensionless constant $g$ for a particular choice of parameters in figure \ref{fig:pheno1}.

The heavy Higgs also decays through the same channels.  The calculation for this rate is the same as the calculation for the light Higgs, with parameters adjusted to match the coefficients in the Lagrangian,
\beq
\mathcal{L} &\supset& \frac{g \lambda_s \mu_w\mu^\gamma}{\sqrt{2}\mup^{1+2\gamma}}v\cos(\alpha-\beta)H^0\mathcal{O} \nonumber \\
 & & -\frac{3\sqrt{2}}{8}\frac{g\lambda_s\mu^\gamma}{\mup^{1+2\gamma}}v\sin(\alpha+\beta)H^0(\partial^2 A+\mathrm{h.c.}).
\eeq

Next we turn to the decay of the LSP of the MSSM, which we will take to be the stau to be concrete.  This decay may be of particular interest, as it has a distinct signature of a single lepton and large missing energy.  Moreover, the kinematics of the decay would reveal the unparticle phase space.  In our model, the stau can decay to a tau and a virtual Higgsino.  The Higgsino then goes to fermionic hidden sector mesons through the derivative couplings in Eq.~(\ref{eq:lha1}).

\begin{figure}[htpb]
\includegraphics[width=8cm]{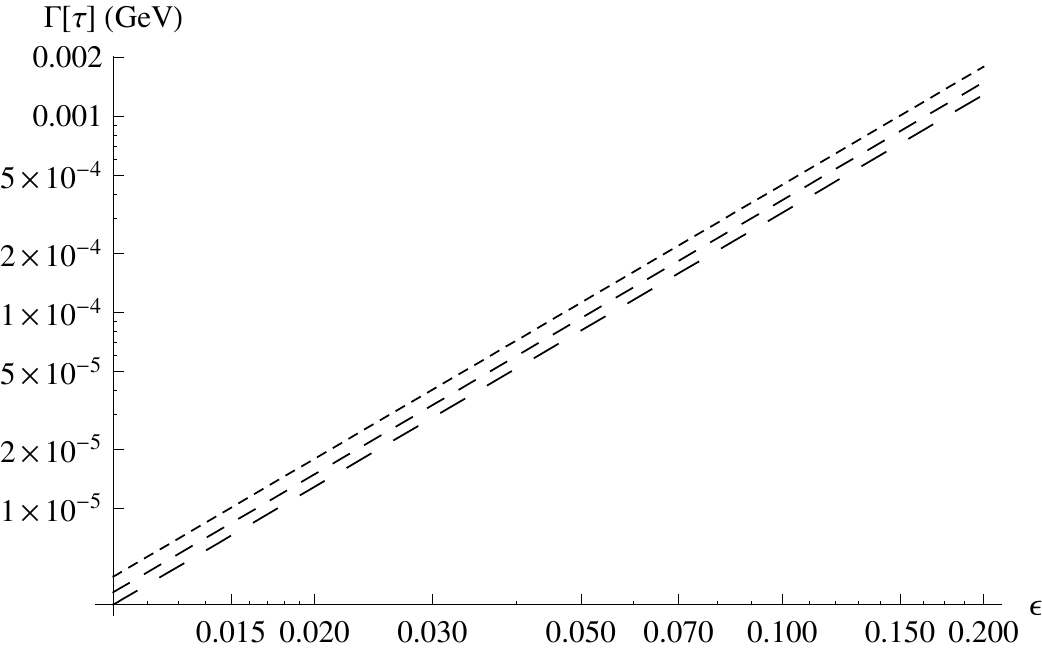}
\caption{The decay rate for the LSP $\tilde{\tau}$ to a $\tau$ and unparticles for the parameters described in the text. As before, the values for $\gamma$ from  shortest to longest dashing are 0.1, 0.3 and 0.5.}
\label{fig:pheno2}
\end{figure}

In general, the momentum structure of phase space does not lend itself to integration.  Closed form solutions exists for certain limits, including the large Higgsino mass case.  In this approximation, the rate is given by
\beq
\Gamma_{\tilde{\tau}\mathrm{dec}} & \approx & c_{\tilde{\tau}} \frac{(m_{\tilde{\tau}}^2-m_\tau^2)^{4+2\gamma}}{m_{\tilde{\tau}}^{3-2\gamma} m_{\tilde{H}}^2(m_{\tilde{\tau}}^2+m_\tau^2)^{2+\gamma}} \nonumber \\
& & \times \left( 2(m_{\tilde{\tau}}^2+m_\tau^2)N_1-(m_{\tilde{\tau}}^2-m_\tau^2)N_2\right)
\eeq
where $N_1$ and $N_2$ are dimensionless numbers of $O(1)$ across most of parameter space, and $M$ is the mass of the LSP.  The overall coupling constant $c_{\tilde{\tau}}$ of dimension $-4\gamma$ is given by
\be
c_{\tilde{\tau}} = \frac{3 (y_\tau g \lambda_s v \sin(\beta))^2}{8\pi^{3/2} \mup^{2+4\gamma}}
\ee
where $y_\tau$ is the $\tau$ Yukawa coupling.
The total decay rate, including the correct numerical factors, is plotted in figure~\ref{fig:pheno2}.  Here we have chosen a Higgsino with mass $m_{\tilde{H}}=800\GeV$, an LSP mass of $m_{\tilde{\tau}}=500\GeV$ and vary $\eps$ as we had for the Higgs decay above.  All other parameters are the same as in the previous cases.

Next we will briefly comment on the relevance for this theory to the widely studied properties of mesons in theories with unparticles~\cite{Aliev:2007gr,Mohanta:2007ad,Lenz:2007nj,Mohanta:2007uu,Aliev:2007rm,Mohanta:2007zq,Wu:2007yh,Aslam:2008th,He:2008xv}.  These papers generally consider the effects of the coupling of the standard model quarks to the unparticle sector
\be
\frac{1}{\Lambda^{d_u-1}} \bar{q}(c_1\gamma_\mu-c_2\gamma_\mu\gamma_5)q\mathcal{O}_U^\mu
\ee
where $c_1$ and $c_2$ are unknown constants.  Here we simply wish to comment that this coupling does not arise at tree level in our theory as the unparticles only couple directly to the Higgs.  Quark-unparticle couplings occurr effectively in penguin diagrams in which the loop quark emits a Higgs, which in turn decays to unparticles through the coupling in \eq{explang}.  These coefficients scale as
\be
c_{1,2}\sim N c_{q_iq_j} \frac{v \lambda_s \mu_w}{\mup^2}
\ee
where $N$ is a small number containing the loop factor and phase space normalization, and $c_{q_iq_j}$ is a value from the CKM matrix that coupes quarks of type $q_i$ and $q_j$ to the Higgs.  The suppression by $\mup$ pushes these coefficients below all established bounds without requiring unnaturally small coefficients.

Finally, we consider the eventual fate of the unparticle stuff that would be produced in the preceding examples.  As discussed previously, below the supersymmetry breaking scale $f_s$ we assume that the unparticle sector becomes confining and forms bound meson and baryon states.  Since there remain strong forces within the hidden sector, we expect that the states which are produced undergo cascade decays and populate stable baryon states, and light meson states.  These latter pseudoscalar particle mix with the pseudoscalar Higgs, $A^0$, which allow them to decay to standard model fermions.  This may produce either interesting topologies or displaced vertices if the pseudoscalar decays occur within the detector.

To estimate the lifetime of these particles, we need to make a small modification to the above assumptions to provide the light states with some mass.  We now assume that all unparticle fields have some small coupling to the Higgs, that is $g_{ii}\ne 0$ for all $i$.  However, we may still have hierarchical couplings so that the processes discussed above are not significantly modified.  We will take the lightest meson state to correspond to $A_{11}$.

The mixing between the hidden pseudoscalar and $A^0$ is given by the first term in \eq{lha1}.  Upon electroweak symmetry break, this gives rise to a coupling
\be
\frac{3\sqrt{2} i \lambda_S g_{11} v \mu^\gamma}{8 \mup^{1+2\gamma}}A^0 \partial^2 A_{11}.
\ee
Integrating out the $A^0$, we find an effective coupling between the hidden meson and the standard model fermions.  

\begin{figure}[htpb]
\includegraphics[width=8cm]{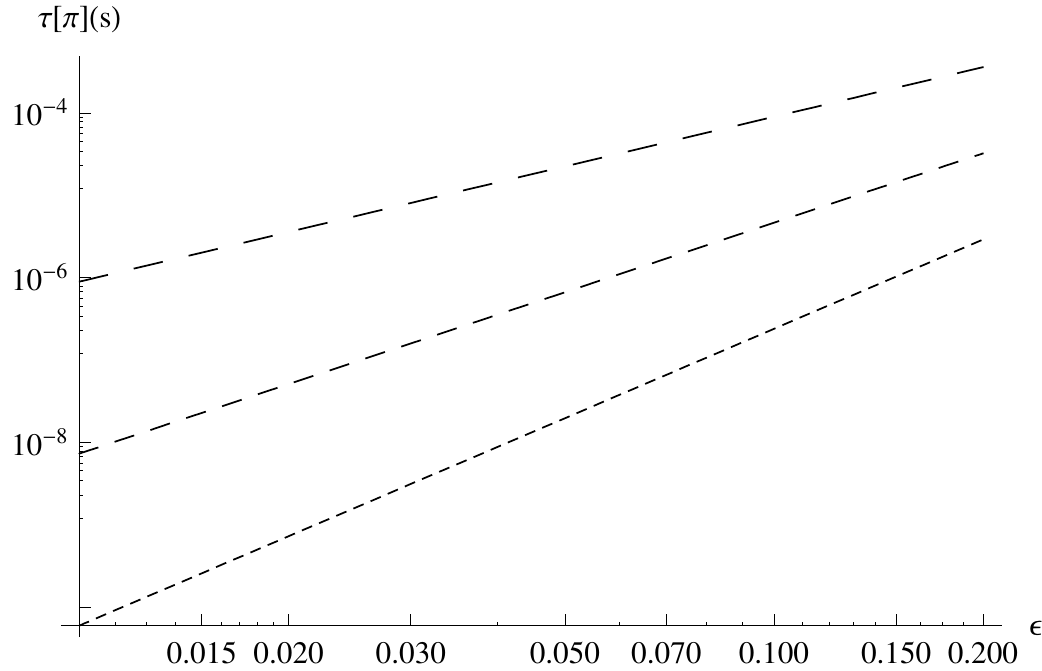}
\caption{The rest frame lifetime for the particle-like lightest hidden meson state, for the simple example discussed in the text, with the overall numerical constant set to 1.  The mass of the state is fixed at $5\GeV$ so that decays are dominantly to taus.  The values for $\gamma$ from  shortest to longest dashing are 0.1, 0.3 and 0.5.}
\label{fig:pheno3}
\end{figure}

We take the mass of the lightest state to be $5\GeV$, just above the $\tau$ production threshold.  The decay to $\tau$s dominates.  There will also be sub-dominant tree-level decays to lighter fermions and loop-supressed decays to gauge bosons that are not considered here.  The effective term that is generated is
\be
c_{\pi_A} \pi_A \bar{\tau} \gamma_5 \tau
\ee
below the scale $f_s$, in which $\pi_A$ is the meson state corresponding to the operator $A_{11}$.  The coupling constant $c_{\pi_A}$ is 
\beq
c_{\pi_A}&=&-C\left( \frac{3\sqrt{2}i\lambda_Sg_{11}vf_s^{2\gamma}}{8\mup^{1+2\gamma}} \frac{m_\tau\tan\beta}{v} \right) \frac{m_{\pi_A}^2}{m_{A^0}^2} \nonumber \\
&=& -C\frac{3\sqrt{2}i\eps_{11}m_\tau\tan\beta}{8f_s} \frac{m_{\pi_A}^2}{m_{A^0}^2}
\eeq
where $\pi_A$ is the meson state corresponding to the operator $A_11$ and $C$ is an unknown constant resulting from the strong dynamics (analogous to the shift in $F_\pi$ from the strong scale in the effective couplings of standard model pions).  Here we have run the coupling down to the scale $f_s$, and included additional powers of $f_s$ to get the correct dimensions.  The resulting rest frame decay rate for the lightest hidden meson is
\be
\Gamma_{\pi_A \mathrm{dec}} = \frac{\mag[c_{\pi_A}]^2 m_{\pi_A}}{8\pi}\left( 1 - \frac{4m_\tau^2}{m_{\pi_A}^2} \right)^\frac{1}{2}.
\ee
To get a sense for the lifetime of these light particles, we plot the rest frame lifetime for the parameters described above in figure \ref{fig:pheno3}.
\section{Summary }
General dimensional analysis arguments indicate that the coupling between the standard model and a light hidden sector   should be dominated by low dimension standard model gauge invariant operators, most likely involving the Higgs. Previous work     has considered couplings between the Higgs and an unparticle sector that induce a gap into the unparticle spectrum. In constrast,
this model illustrates a different possibility, where the coupling between the Higgs and the unparticle sector drives the unparticle sector to a new   conformal fixed point in the infrared. We illustrate how to couple the unparticle sector to the Higgs through an effective Giudice-Masiero mechanism, which   shields the unparticles from the effects of supersymmetry breaking, so that the supersymmetry breaking scale is low and supersymmetry may be used for theoretical control over the computations.  Below the supersymmetry breaking scale we must make dynamical assumptions about the  strongly coupled theory. We assume the very  low energy theory is confining.

There are several ways  in which the hidden sector may be produced and observed in colliders. The Higgs may dominantly decay into unparticles. The exotic kinematics of unparticle decays are conceivably   observable through the decays of the lightest MSSM superpartner into a visible particle and unparticles.
Once the unparticles are produced, at long distances they hadronize. There are a multitude of stable and meta stable particles with a broad spectrum of masses in the unparticle sector, providing interesting possibilities for dark matter.  As  in ``hidden valley'' models of strongly coupled hidden sectors, some of the  hidden hadrons decay promptly, some with displaced vertices, and some are stable. 
These models require further study to understand  their implications  for detection of the Higgs, supersymmetry and dark matter.

\noindent {\bf Acknowledgments}

  This work was
 supported in part by DOE grant DE-FGO3-00ER41132.
 The work of MP is supported in part  by the Wales Institute of
Mathematical and Computational Sciences, and by the STFC 
grant ST/G000506/1.

\bibliography{unparticle}

\begin{thebibliography}{40}
\expandafter\ifx\csname natexlab\endcsname\relax\def\natexlab#1{#1}\fi
\expandafter\ifx\csname bibnamefont\endcsname\relax
  \def\bibnamefont#1{#1}\fi
\expandafter\ifx\csname bibfnamefont\endcsname\relax
  \def\bibfnamefont#1{#1}\fi
\expandafter\ifx\csname citenamefont\endcsname\relax
  \def\citenamefont#1{#1}\fi
\expandafter\ifx\csname url\endcsname\relax
  \def\url#1{\texttt{#1}}\fi
\expandafter\ifx\csname urlprefix\endcsname\relax\def\urlprefix{URL }\fi
\providecommand{\bibinfo}[2]{#2}
\providecommand{\eprint}[2][]{\url{#2}}

\bibitem[{\citenamefont{Strassler and Zurek}(2007)}]{Strassler:2006im}
\bibinfo{author}{\bibfnamefont{M.~J.} \bibnamefont{Strassler}}
  \bibnamefont{and} \bibinfo{author}{\bibfnamefont{K.~M.} \bibnamefont{Zurek}},
  \bibinfo{journal}{Phys. Lett.} \textbf{\bibinfo{volume}{B651}},
  \bibinfo{pages}{374} (\bibinfo{year}{2007}), \eprint{hep-ph/0604261}.

\bibitem[{\citenamefont{Strassler and Zurek}(2008)}]{Strassler:2006ri}
\bibinfo{author}{\bibfnamefont{M.~J.} \bibnamefont{Strassler}}
  \bibnamefont{and} \bibinfo{author}{\bibfnamefont{K.~M.} \bibnamefont{Zurek}},
  \bibinfo{journal}{Phys. Lett.} \textbf{\bibinfo{volume}{B661}},
  \bibinfo{pages}{263} (\bibinfo{year}{2008}), \eprint{hep-ph/0605193}.

\bibitem[{\citenamefont{Georgi}(2007{\natexlab{a}})}]{Georgi:2007ek}
\bibinfo{author}{\bibfnamefont{H.}~\bibnamefont{Georgi}},
  \bibinfo{journal}{Phys. Rev. Lett.} \textbf{\bibinfo{volume}{98}},
  \bibinfo{pages}{221601} (\bibinfo{year}{2007}{\natexlab{a}}),
  \eprint{hep-ph/0703260}.

\bibitem[{\citenamefont{Georgi}(2007{\natexlab{b}})}]{Georgi:2007si}
\bibinfo{author}{\bibfnamefont{H.}~\bibnamefont{Georgi}},
  \bibinfo{journal}{Phys. Lett.} \textbf{\bibinfo{volume}{B650}},
  \bibinfo{pages}{275} (\bibinfo{year}{2007}{\natexlab{b}}),
  \eprint{0704.2457}.

\bibitem[{\citenamefont{Stephanov}(2007)}]{Stephanov:2007ry}
\bibinfo{author}{\bibfnamefont{M.~A.} \bibnamefont{Stephanov}},
  \bibinfo{journal}{Phys. Rev.} \textbf{\bibinfo{volume}{D76}},
  \bibinfo{pages}{035008} (\bibinfo{year}{2007}), \eprint{0705.3049}.

\bibitem[{\citenamefont{Cacciapaglia et~al.}(2009)\citenamefont{Cacciapaglia,
  Marandella, and Terning}}]{Cacciapaglia:2008ns}
\bibinfo{author}{\bibfnamefont{G.}~\bibnamefont{Cacciapaglia}},
  \bibinfo{author}{\bibfnamefont{G.}~\bibnamefont{Marandella}},
  \bibnamefont{and} \bibinfo{author}{\bibfnamefont{J.}~\bibnamefont{Terning}},
  \bibinfo{journal}{JHEP} \textbf{\bibinfo{volume}{02}}, \bibinfo{pages}{049}
  (\bibinfo{year}{2009}), \eprint{0804.0424}.

\bibitem[{\citenamefont{Strassler}(2008)}]{Strassler:2008bv}
\bibinfo{author}{\bibfnamefont{M.~J.} \bibnamefont{Strassler}}
  (\bibinfo{year}{2008}), \eprint{0801.0629}.

\bibitem[{\citenamefont{Sannino and Zwicky}(2009)}]{Sannino:2008nv}
\bibinfo{author}{\bibfnamefont{F.}~\bibnamefont{Sannino}} \bibnamefont{and}
  \bibinfo{author}{\bibfnamefont{R.}~\bibnamefont{Zwicky}},
  \bibinfo{journal}{Phys. Rev.} \textbf{\bibinfo{volume}{D79}},
  \bibinfo{pages}{015016} (\bibinfo{year}{2009}), \eprint{0810.2686}.

\bibitem[{\citenamefont{Banks and Zaks}(1982)}]{Banks:1981nn}
\bibinfo{author}{\bibfnamefont{T.}~\bibnamefont{Banks}} \bibnamefont{and}
  \bibinfo{author}{\bibfnamefont{A.}~\bibnamefont{Zaks}},
  \bibinfo{journal}{Nucl. Phys.} \textbf{\bibinfo{volume}{B196}},
  \bibinfo{pages}{189} (\bibinfo{year}{1982}).

\bibitem[{\citenamefont{Seiberg}(1995)}]{Seiberg:1994pq}
\bibinfo{author}{\bibfnamefont{N.}~\bibnamefont{Seiberg}},
  \bibinfo{journal}{Nucl. Phys.} \textbf{\bibinfo{volume}{B435}},
  \bibinfo{pages}{129} (\bibinfo{year}{1995}), \eprint{hep-th/9411149}.

\bibitem[{\citenamefont{Intriligator and Seiberg}(1996)}]{Intriligator:1995au}
\bibinfo{author}{\bibfnamefont{K.~A.} \bibnamefont{Intriligator}}
  \bibnamefont{and} \bibinfo{author}{\bibfnamefont{N.}~\bibnamefont{Seiberg}},
  \bibinfo{journal}{Nucl. Phys. Proc. Suppl.} \textbf{\bibinfo{volume}{45BC}},
  \bibinfo{pages}{1} (\bibinfo{year}{1996}), \eprint{hep-th/9509066}.

\bibitem[{\citenamefont{Fox et~al.}(2007)\citenamefont{Fox, Rajaraman, and
  Shirman}}]{Fox:2007sy}
\bibinfo{author}{\bibfnamefont{P.~J.} \bibnamefont{Fox}},
  \bibinfo{author}{\bibfnamefont{A.}~\bibnamefont{Rajaraman}},
  \bibnamefont{and} \bibinfo{author}{\bibfnamefont{Y.}~\bibnamefont{Shirman}},
  \bibinfo{journal}{Phys. Rev.} \textbf{\bibinfo{volume}{D76}},
  \bibinfo{pages}{075004} (\bibinfo{year}{2007}), \eprint{0705.3092}.

\bibitem[{\citenamefont{Delgado et~al.}(2007)\citenamefont{Delgado, Espinosa,
  and Quiros}}]{Delgado:2007dx}
\bibinfo{author}{\bibfnamefont{A.}~\bibnamefont{Delgado}},
  \bibinfo{author}{\bibfnamefont{J.~R.} \bibnamefont{Espinosa}},
  \bibnamefont{and} \bibinfo{author}{\bibfnamefont{M.}~\bibnamefont{Quiros}},
  \bibinfo{journal}{JHEP} \textbf{\bibinfo{volume}{10}}, \bibinfo{pages}{094}
  (\bibinfo{year}{2007}), \eprint{0707.4309}.

\bibitem[{\citenamefont{Kikuchi and Okada}(2008)}]{Kikuchi:2007qd}
\bibinfo{author}{\bibfnamefont{T.}~\bibnamefont{Kikuchi}} \bibnamefont{and}
  \bibinfo{author}{\bibfnamefont{N.}~\bibnamefont{Okada}},
  \bibinfo{journal}{Phys. Lett.} \textbf{\bibinfo{volume}{B661}},
  \bibinfo{pages}{360} (\bibinfo{year}{2008}), \eprint{0707.0893}.

\bibitem[{\citenamefont{Dine and Nelson}(1993)}]{Dine:1993yw}
\bibinfo{author}{\bibfnamefont{M.}~\bibnamefont{Dine}} \bibnamefont{and}
  \bibinfo{author}{\bibfnamefont{A.~E.} \bibnamefont{Nelson}},
  \bibinfo{journal}{Phys. Rev.} \textbf{\bibinfo{volume}{D48}},
  \bibinfo{pages}{1277} (\bibinfo{year}{1993}), \eprint{hep-ph/9303230}.

\bibitem[{\citenamefont{Dine et~al.}(1995)\citenamefont{Dine, Nelson, and
  Shirman}}]{Dine:1994vc}
\bibinfo{author}{\bibfnamefont{M.}~\bibnamefont{Dine}},
  \bibinfo{author}{\bibfnamefont{A.~E.} \bibnamefont{Nelson}},
  \bibnamefont{and} \bibinfo{author}{\bibfnamefont{Y.}~\bibnamefont{Shirman}},
  \bibinfo{journal}{Phys. Rev.} \textbf{\bibinfo{volume}{D51}},
  \bibinfo{pages}{1362} (\bibinfo{year}{1995}), \eprint{hep-ph/9408384}.

\bibitem[{\citenamefont{Dine et~al.}(1996)\citenamefont{Dine, Nelson, Nir, and
  Shirman}}]{Dine:1995ag}
\bibinfo{author}{\bibfnamefont{M.}~\bibnamefont{Dine}},
  \bibinfo{author}{\bibfnamefont{A.~E.} \bibnamefont{Nelson}},
  \bibinfo{author}{\bibfnamefont{Y.}~\bibnamefont{Nir}}, \bibnamefont{and}
  \bibinfo{author}{\bibfnamefont{Y.}~\bibnamefont{Shirman}},
  \bibinfo{journal}{Phys. Rev.} \textbf{\bibinfo{volume}{D53}},
  \bibinfo{pages}{2658} (\bibinfo{year}{1996}), \eprint{hep-ph/9507378}.

\bibitem[{\citenamefont{Giudice and Masiero}(1988)}]{Giudice:1988yz}
\bibinfo{author}{\bibfnamefont{G.~F.} \bibnamefont{Giudice}} \bibnamefont{and}
  \bibinfo{author}{\bibfnamefont{A.}~\bibnamefont{Masiero}},
  \bibinfo{journal}{Phys. Lett.} \textbf{\bibinfo{volume}{B206}},
  \bibinfo{pages}{480} (\bibinfo{year}{1988}).

\bibitem[{\citenamefont{Nelson and Strassler}(2002)}]{Nelson:2001mq}
\bibinfo{author}{\bibfnamefont{A.~E.} \bibnamefont{Nelson}} \bibnamefont{and}
  \bibinfo{author}{\bibfnamefont{M.~J.} \bibnamefont{Strassler}},
  \bibinfo{journal}{JHEP} \textbf{\bibinfo{volume}{07}}, \bibinfo{pages}{021}
  (\bibinfo{year}{2002}), \eprint{hep-ph/0104051}.

\bibitem[{\citenamefont{Nelson and Spitzer}(2008)}]{Nelson:2008hj}
\bibinfo{author}{\bibfnamefont{A.~E.} \bibnamefont{Nelson}} \bibnamefont{and}
  \bibinfo{author}{\bibfnamefont{C.}~\bibnamefont{Spitzer}}
  (\bibinfo{year}{2008}), \eprint{0810.5167}.

\bibitem[{\citenamefont{Adriani et~al.}(2009)}]{Adriani:2008zr}
\bibinfo{author}{\bibfnamefont{O.}~\bibnamefont{Adriani}} \bibnamefont{et~al.}
  (\bibinfo{collaboration}{PAMELA}), \bibinfo{journal}{Nature}
  \textbf{\bibinfo{volume}{458}}, \bibinfo{pages}{607} (\bibinfo{year}{2009}),
  \eprint{0810.4995}.

\bibitem[{\citenamefont{Tucker-Smith and Weiner}(2005)}]{TuckerSmith:2004jv}
\bibinfo{author}{\bibfnamefont{D.}~\bibnamefont{Tucker-Smith}}
  \bibnamefont{and} \bibinfo{author}{\bibfnamefont{N.}~\bibnamefont{Weiner}},
  \bibinfo{journal}{Phys. Rev.} \textbf{\bibinfo{volume}{D72}},
  \bibinfo{pages}{063509} (\bibinfo{year}{2005}), \eprint{hep-ph/0402065}.

\bibitem[{\citenamefont{Bernabei et~al.}(2008)}]{Bernabei:2008yi}
\bibinfo{author}{\bibfnamefont{R.}~\bibnamefont{Bernabei}} \bibnamefont{et~al.}
  (\bibinfo{collaboration}{DAMA}), \bibinfo{journal}{Eur. Phys. J.}
  \textbf{\bibinfo{volume}{C56}}, \bibinfo{pages}{333} (\bibinfo{year}{2008}),
  \eprint{0804.2741}.

\bibitem[{\citenamefont{Finkbeiner and Weiner}(2007)}]{Finkbeiner:2007kk}
\bibinfo{author}{\bibfnamefont{D.~P.} \bibnamefont{Finkbeiner}}
  \bibnamefont{and} \bibinfo{author}{\bibfnamefont{N.}~\bibnamefont{Weiner}},
  \bibinfo{journal}{Phys. Rev.} \textbf{\bibinfo{volume}{D76}},
  \bibinfo{pages}{083519} (\bibinfo{year}{2007}), \eprint{astro-ph/0702587}.

\bibitem[{\citenamefont{Knodlseder et~al.}(2005)}]{Knodlseder:2005yq}
\bibinfo{author}{\bibfnamefont{J.}~\bibnamefont{Knodlseder}}
  \bibnamefont{et~al.}, \bibinfo{journal}{Astron. Astrophys.}
  \textbf{\bibinfo{volume}{441}}, \bibinfo{pages}{513} (\bibinfo{year}{2005}),
  \eprint{astro-ph/0506026}.

\bibitem[{\citenamefont{Bander et~al.}(2007)\citenamefont{Bander, Feng,
  Rajaraman, and Shirman}}]{Bander:2007nd}
\bibinfo{author}{\bibfnamefont{M.}~\bibnamefont{Bander}},
  \bibinfo{author}{\bibfnamefont{J.~L.} \bibnamefont{Feng}},
  \bibinfo{author}{\bibfnamefont{A.}~\bibnamefont{Rajaraman}},
  \bibnamefont{and} \bibinfo{author}{\bibfnamefont{Y.}~\bibnamefont{Shirman}},
  \bibinfo{journal}{Phys. Rev.} \textbf{\bibinfo{volume}{D76}},
  \bibinfo{pages}{115002} (\bibinfo{year}{2007}), \eprint{0706.2677}.

\bibitem[{\citenamefont{Luo and Zhu}(2008)}]{Luo:2007bq}
\bibinfo{author}{\bibfnamefont{M.}~\bibnamefont{Luo}} \bibnamefont{and}
  \bibinfo{author}{\bibfnamefont{G.}~\bibnamefont{Zhu}},
  \bibinfo{journal}{Phys. Lett.} \textbf{\bibinfo{volume}{B659}},
  \bibinfo{pages}{341} (\bibinfo{year}{2008}), \eprint{0704.3532}.

\bibitem[{\citenamefont{Basu et~al.}(2008)\citenamefont{Basu, Choudhury, and
  Mani}}]{Basu:2008rd}
\bibinfo{author}{\bibfnamefont{R.}~\bibnamefont{Basu}},
  \bibinfo{author}{\bibfnamefont{D.}~\bibnamefont{Choudhury}},
  \bibnamefont{and} \bibinfo{author}{\bibfnamefont{H.~S.} \bibnamefont{Mani}}
  (\bibinfo{year}{2008}), \eprint{0803.4110}.

\bibitem[{\citenamefont{Liao}(2008)}]{Liao:2008tj}
\bibinfo{author}{\bibfnamefont{Y.}~\bibnamefont{Liao}}, \bibinfo{journal}{Phys.
  Lett.} \textbf{\bibinfo{volume}{B665}}, \bibinfo{pages}{356}
  (\bibinfo{year}{2008}), \eprint{0804.0752}.

\bibitem[{\citenamefont{Delgado et~al.}(2009)\citenamefont{Delgado, Espinosa,
  No, and Quiros}}]{Delgado:2008gj}
\bibinfo{author}{\bibfnamefont{A.}~\bibnamefont{Delgado}},
  \bibinfo{author}{\bibfnamefont{J.~R.} \bibnamefont{Espinosa}},
  \bibinfo{author}{\bibfnamefont{J.~M.} \bibnamefont{No}}, \bibnamefont{and}
  \bibinfo{author}{\bibfnamefont{M.}~\bibnamefont{Quiros}},
  \bibinfo{journal}{Phys. Rev.} \textbf{\bibinfo{volume}{D79}},
  \bibinfo{pages}{055011} (\bibinfo{year}{2009}), \eprint{0812.1170}.

\bibitem[{\citenamefont{Chang et~al.}(2008)\citenamefont{Chang, Dermisek,
  Gunion, and Weiner}}]{Chang:2008cw}
\bibinfo{author}{\bibfnamefont{S.}~\bibnamefont{Chang}},
  \bibinfo{author}{\bibfnamefont{R.}~\bibnamefont{Dermisek}},
  \bibinfo{author}{\bibfnamefont{J.~F.} \bibnamefont{Gunion}},
  \bibnamefont{and} \bibinfo{author}{\bibfnamefont{N.}~\bibnamefont{Weiner}},
  \bibinfo{journal}{Ann. Rev. Nucl. Part. Sci.} \textbf{\bibinfo{volume}{58}},
  \bibinfo{pages}{75} (\bibinfo{year}{2008}), \eprint{0801.4554}.

\bibitem[{\citenamefont{Aliev et~al.}(2007)\citenamefont{Aliev, Cornell, and
  Gaur}}]{Aliev:2007gr}
\bibinfo{author}{\bibfnamefont{T.~M.} \bibnamefont{Aliev}},
  \bibinfo{author}{\bibfnamefont{A.~S.} \bibnamefont{Cornell}},
  \bibnamefont{and} \bibinfo{author}{\bibfnamefont{N.}~\bibnamefont{Gaur}},
  \bibinfo{journal}{JHEP} \textbf{\bibinfo{volume}{07}}, \bibinfo{pages}{072}
  (\bibinfo{year}{2007}), \eprint{0705.4542}.

\bibitem[{\citenamefont{Mohanta and Giri}(2007{\natexlab{a}})}]{Mohanta:2007ad}
\bibinfo{author}{\bibfnamefont{R.}~\bibnamefont{Mohanta}} \bibnamefont{and}
  \bibinfo{author}{\bibfnamefont{A.~K.} \bibnamefont{Giri}},
  \bibinfo{journal}{Phys. Rev.} \textbf{\bibinfo{volume}{D76}},
  \bibinfo{pages}{075015} (\bibinfo{year}{2007}{\natexlab{a}}),
  \eprint{0707.1234}.

\bibitem[{\citenamefont{Lenz}(2007)}]{Lenz:2007nj}
\bibinfo{author}{\bibfnamefont{A.}~\bibnamefont{Lenz}}, \bibinfo{journal}{Phys.
  Rev.} \textbf{\bibinfo{volume}{D76}}, \bibinfo{pages}{065006}
  (\bibinfo{year}{2007}), \eprint{0707.1535}.

\bibitem[{\citenamefont{Mohanta and Giri}(2007{\natexlab{b}})}]{Mohanta:2007uu}
\bibinfo{author}{\bibfnamefont{R.}~\bibnamefont{Mohanta}} \bibnamefont{and}
  \bibinfo{author}{\bibfnamefont{A.~K.} \bibnamefont{Giri}},
  \bibinfo{journal}{Phys. Rev.} \textbf{\bibinfo{volume}{D76}},
  \bibinfo{pages}{057701} (\bibinfo{year}{2007}{\natexlab{b}}),
  \eprint{0707.3308}.

\bibitem[{\citenamefont{Aliev and Savci}(2008)}]{Aliev:2007rm}
\bibinfo{author}{\bibfnamefont{T.~M.} \bibnamefont{Aliev}} \bibnamefont{and}
  \bibinfo{author}{\bibfnamefont{M.}~\bibnamefont{Savci}},
  \bibinfo{journal}{Phys. Lett.} \textbf{\bibinfo{volume}{B662}},
  \bibinfo{pages}{165} (\bibinfo{year}{2008}), \eprint{0710.1505}.

\bibitem[{\citenamefont{Mohanta and Giri}(2008)}]{Mohanta:2007zq}
\bibinfo{author}{\bibfnamefont{R.}~\bibnamefont{Mohanta}} \bibnamefont{and}
  \bibinfo{author}{\bibfnamefont{A.~K.} \bibnamefont{Giri}},
  \bibinfo{journal}{Phys. Lett.} \textbf{\bibinfo{volume}{B660}},
  \bibinfo{pages}{376} (\bibinfo{year}{2008}), \eprint{0711.3516}.

\bibitem[{\citenamefont{Wu and Zhang}(2007)}]{Wu:2007yh}
\bibinfo{author}{\bibfnamefont{Y.-f.} \bibnamefont{Wu}} \bibnamefont{and}
  \bibinfo{author}{\bibfnamefont{D.-X.} \bibnamefont{Zhang}}
  (\bibinfo{year}{2007}), \eprint{0712.3923}.

\bibitem[{\citenamefont{Aslam and Lu}(2008)}]{Aslam:2008th}
\bibinfo{author}{\bibfnamefont{M.~J.} \bibnamefont{Aslam}} \bibnamefont{and}
  \bibinfo{author}{\bibfnamefont{C.-D.} \bibnamefont{Lu}}
  (\bibinfo{year}{2008}), \eprint{0802.0739}.

\bibitem[{\citenamefont{He and Tsai}(2008)}]{He:2008xv}
\bibinfo{author}{\bibfnamefont{X.-G.} \bibnamefont{He}} \bibnamefont{and}
  \bibinfo{author}{\bibfnamefont{L.}~\bibnamefont{Tsai}},
  \bibinfo{journal}{JHEP} \textbf{\bibinfo{volume}{06}}, \bibinfo{pages}{074}
  (\bibinfo{year}{2008}), \eprint{0805.3020}.

\end{thebibliography}
\bibliographystyle{apsrev}

\section*{Appendix: large-$N$ counting.}

In order to provide a  set of estimates for the 
masses of the particles that decouple from the unparticle sector in the 
hidden sector when symmetry breaking takes place 
(at the scale $\Lambda$), it is useful to consider the limiting case in which $\gamma\simeq 0$,
so that perturbative techniques can be used, and
study the large-$N$ behavior of the physical quantities.

The masses of the heavy vector superfields resulting from 
symmetry-breaking in the magnetic sector are
\beqs
m_V^2&\simeq&\frac{1}{2}g_s^2m^2\,,
\eeqs
where 
\beqs
m^2&=&\langle \bar{q}_iq_i\rangle\,.
\eeqs

In order for these masses to be finite, one has to
assume that 
\beqs
m^2&\propto&N_c\,,
\eeqs
since the gauge coupling itself $g_s^2\propto 1/N_c$.

The masses of the  chiral (meson) superfields 
corresponding to the pseudo-Goldstone bosons 
carrying one unbroken flavor index $I$ and one index along the broken directions $N$
are
\beqs
m^2_{A_{NI}}&\simeq&\lambda^2 m^2\,,
\eeqs
which implies that 
\beqs
\lambda^2&\propto&\frac{1}{N_c}\,.
\eeqs

As a consequence, the scaling of $m^2$ requires 
 that $\epsilon_{ii}/\lambda\propto N_c$, 
 and hence $\epsilon_{ii} \propto g_{ii} \propto \sqrt{N_c}$.
The relevant
couplings $g_{ii}$ grow with $N_c$ as
\beqs
g_{ii}^2&\propto& N_c\,.
\eeqs

The  massive meson superfields carrying only indexes on the 
broken directions $A^{\prime}_{N N}$
have mass
\beqs
m^2_{A_{N N}}&\simeq&
\frac{2\lambda^2m^2}{1+g_{ii}^2}
\,.
\eeqs
This is suppressed in the large-$N_c$ limit.

Also the mass of  linear combination of quarks and 
anti-quarks carrying both flavor and color indexes in the broken directions 
is suppressed as 
$m^2_{q_N}\simeq m^2_{A_{N N}} \propto 1/N_c$.

This fixes the large-$N_c$ scaling of all the parameters and resulting masses.
Notice how the same scalings would have been obtained by 
rescaling $A\rightarrow \tilde{A}/g$ for all the magnetic fields, but not
for the MSSM and messenger fields, and requiring that all terms
we wrote in the superpotential appear at the same order in the $1/N$ expansion.
Interestingly, besides expected results, we find that in the large-$N_c$ limit the 
fields corresponding to perturbations in the transverse directions
of the potential become light.

All of these expressions generalize to the case where $\gamma>0$
by replacing $g_{ii}\rightarrow g_{ii}(v/\mu^{\prime})^{\gamma}$, while $m^2$
has to be evaluated at the electro-weak scale (as done in the body of the paper).
The couplings $\lambda$ and $g_s$ do not run between $v$ and $\mu^{\prime}$.

\end{document}